# Wheelchair Automation by a Hybrid BCI System Using SSVEP and Eye Blinks

Lizy Kanungo, Nikhil Garg, Anish Bhobe, Smit Rajguru, Veeky Baths

*Abstract*— This work proposes a hybrid Brain-Computer Interface system for the automation of a wheelchair for the disabled. Herein a working prototype of a BCI-based wheelchair is detailed that can navigate inside a typical home environment with minimum structural modification and without any visual obstruction and discomfort to the user. The prototype is based on a combined mechanism of steady-state visually evoked potential and eye blinks. To elicit SSVEP, LEDs flickering at 13Hz and 15Hz were used to select the left and right direction, respectively, and EEG data was recorded. In addition, the occurrence of three continuous blinks was used as an indicator for stopping an ongoing action. The wavelet packet denoising method was applied, followed by feature extraction methods such as Wavelet Packet Decomposition and Canonical Correlation Analysis over narrowband reconstructed EEG signals. Bayesian optimization was used to obtain 5-fold cross-validations to optimize the hyperparameters of the Support Vector Machine. The resulting new model was tested and the average cross-validation accuracy 89.65% $\pm$ 6.6% (SD) and testing accuracy 83.53% $\pm$ 8.59% (SD) were obtained. The wheelchair was controlled by RaspberryPi through WiFi. The developed prototype demonstrated an average of 86.97% success rate for all trials with 4.015s for each command execution. The prototype can be used efficiently in a home environment without causing any discomfort to the user.

## I. Introduction

Electroencephalography or EEG is a non-invasive electrophysiological monitoring method used to capture brain signals. As EEG has a very high temporal resolution over other techniques such as fMRI, it is most widely used for Brain Computer Interface (BCI) that connects the computer directly to the human brain as the operator. EEG-based BCI has been quite popular due to its simple and safe approach [1]. It is also quite useful for the full-body paralyzed people who cannot move their limbs to execute simple tasks. Some of the brain activities that can be effectively recorded from the scalp by using EEG Slow Cortical Potentials (SCPs), P300 potentials and Steady state visually evoked potential (SSVEP) [2]. Among them, SSVEPs have caught attention due to their advantages of requiring less or no training, high Information Transfer Rate (ITR) and ease of use [3-4]. Fundamentally, SSVEPs are elicited responses in the human brain when a person visually focuses his/her attention on a Repetitive Visual Stimulus (RVS), a source flickering at frequency 6Hz or above [5].

The resulting signals can be recorded via an EEG system. These signals are strong in the occipital region of the brain and are sinusoidal waveforms. These waveforms have the same fundamental frequency as the stimulus, including some of its harmonics. By matching the fundamental frequency of the SSVEP to one of the stimulus frequencies presented, it is possible to detect the target selected by the user [6]. SSVEP has been widely used in many home automation researches due to its simplicity and accuracy [2-5]. In SSVEP based BCI research, the major work cited in the literature is based on Graphical User Interface (GUI) [7-10]. In current BCI methods, facilitating the multi-direction-oriented movements are very much in demand for the physically disabled persons who are wheelchair dependent and are devoid of any limb movement for their communication. The idea is to provide them with ease of movement without any external support. However, practically, the SSVEP-based wheelchair automation suffers from several difficulties, such as vision obstruction due to the frontal placement of the LED screen. Usually, a minimum of 4 LEDs are required for navigation in 4 directions, such as right, left, start and stop. This creates a requirement of the 4 class BCI system. Traditional SSVEP based wheelchair requires an LCD panel to elicit 4 different frequencies. The position of the stimuli in such SSVEP-based wheelchairs is often fixed. This creates visual obstruction and restricts freedom of movement and field of vision.

Moreover, the user must concentrate on the LCD all the time for front movement, which restricts the eye movement to observe the environment for safe navigation. While the user is navigating through a wheelchair, his/her eyes can occasionally move away from the stimuli. Furthermore, the user can be distracted, which can deteriorate the signal because the SSVEP strength is strongly influenced by attention. Moreover, the LED intended for forward movement often causes strain on the eyes due to continuous staring.

We have summarized the SSVEP based wheelchair prototypes reported in the literature in Table 1. All of the reported studies considered SSVEP protocol for the forward and backward movement. Moreover, an LCD screen was mounted over the wheelchair to generate the stimulus, which leads to high cost and discomfort to the user as the field of vision is limited.

* This work was funded by the Department of Science and Technology, Government of India, vide Reference No: SR/CSI/50/2014 (G) through the Cognitive Science Research Initiative (CSRI). We thank BITS, Pilani-K.K. Birla Goa Campus for research aid.

Lizy Kanungo, Nikhil Garg, Anish Bhobe, Smit Rajguru, Veeky Baths are with the Cognitive Neuroscience Lab, Department of Biological Science, BITS, Pilani- K. K. Birla Goa Campus, Goa, India 403726 (e-mail: lizy_kanungo@yahoo.co.in).

To address the above issue, BCI commands such as Eyeblink can give freedom of visual attention during navigation. However, for turning right/left, SSVEP commands can be used that are produced by LEDs placed in respective directions over the armrest of the wheelchair. The main advantage of this proposed BCI is a low-cost and straightforward setup as sophisticated LCD screens and mounting arrangement is not needed for stimulus with little effort required to adjust the system parameters to the user. Herein we have used the novel combination of SSVEP and eye blink signals to control our electric wheelchair prototype. When the number of stimuli is increased and flickering with its frequency, it is possible to determine which of the stimuli was observed by a subject. We have used 2 LEDs flickering at 13 and 15 Hz for Left/Right direction control in our prototype. Since eye blink is one of the most straightforward tasks/mechanisms that can help disabled people to do their everyday routines, we have associated them with the GO and STOP command. Additionally, it is equipped with ultrasonic sensors for obstacle detection and an emergency force stop system to avoid any possible collision or injury to the user.

TABLE I
COMPARATIVE ANALYSIS OF DIFFERENT SSVEP TECHNIQUES USED FOR WHEELCHAIR MOVEMENT.

| Sl. No | Frequency used (Hz) | Feature extraction methods | ML used | No. of subjects | Accuracy obtained | Reference |
|---|---|---|---|---|---|---|
| 1 | 37, 38, 39, 40 | Power Spectral Density (PSD) | Statistical maximum | 15 | 44.6 bits/min | [11] |
| 2 | 15, 12, 10, 8.5 | FFT and CCA | CCA coefficient | 4 | 95.25 | [12] |
| 3 | 11,12,13, 14, 15 | Amplitude of the fundamental frequency | Threshold method not specified. | 37 | >50% | [13] |
| 4 | 5.6, 6.4, 6.9, 8 | Frequency peaks | Decision tree method | 9 | precision of 83 ± 15 % | [14] |
| 5 | 9, 10, 11,12 | CCA | Bayesian | 5 | 87.17% | [15] |
| 6 | 13,14,15,16 | Frequency band power | Threshold method not specified | 9 | 88.20% | [16] |
| 7 | 13, 15 | CCA over reconstructed WPD nodes. | SVM | 12 | 83.53%± 8.59 (SD) | This work |

## II. MATERIALS AND METHODOLOGIES

### A. Subjects

12 healthy volunteers (11 male, 1 female) aged between 20-24 (average age 22) with normal or corrected-to-normal vision participated in the study. Among them, 11 were right-handed, and only 1 was left-handed. The subjects were student volunteers with no history of neurological or psychiatric diseases. All were informed about the testing procedure and signed written consent. The subjects have also given their consent for publication of identifying information/images in an online open-access publication if needed. A full explanation of the experiment was provided to each of the subjects. The study was approved by the Institutional Ethics Committee of BITS, Pilani (IHEC-40/16-1). All EEG experiments/methods were performed in accordance with the relevant guidelines and regulations as per EGI, Inc (Electrical Geodesics Inc., Eugene, OR, USA) and the Institutional Ethics Committee of BITS, Pilani.

### B. Experimental setup

#### 1) SSVEP detection

Two red LEDs, 13Hz and 15Hz, were selected for our study as mid-range frequencies were reported to be efficient and user-friendly in many SSVEP studies [17-19]. Though it is reported [20,21] that red colour in mid-range frequencies may cause epileptic seizures and fatigue in some people who stared at it for a long time, it has the highest ITR compared to other colours [22,23]. The intention of using it here is that it is intended for only left/right direction selection commands which do not demand more time in practical use (limited to a maximum of 4s). Therefore, the chance of inducing fatigue was avoided. The tiny LED bulbs (3mm; c/v: 20mA, 2v-2.2v) were placed in front of the subject at approx. 60cm, mounted on the left and right armrest (13Hz and 15Hz respectively) of the chair. An Arduino device controlled the LEDs. The EEG was recorded (Ag/AgCl electrodes, initially referenced to the vertex) using a 32-channel Electrical Geodesics sensor net (Electrical Geodesics Inc., Eugene, OR, USA). The electrodes were placed as per the 10-20 International positioning system. The impedances were maintained below 50KΩ as per the EGI guidelines. Each experiment consisted of a calibration session followed by an online testing session.

The subjects were asked to concentrate on the left LED, right LED, or straight by audio commands. Each trial lasted for 15s. For each participant, 10 trials for each LED and Baseline were obtained. Signals were sampled at 1000 Hz. Bad channels were identified with visual inspection and were removed. Among the 32 electrodes, O1, O2, and Oz (9, 10, 20) electrodes were selected and used for analysis as they were most relevant to our experiments. The data were analyzed using MATLAB™, and the classifier was trained. Subsequently, online sessions were carried out where the subjects controlled the wheelchair by concentrating on intended LEDs and blinks. Each such trial was of 4s duration. For each subject, 20 trials in each left and right direction were carried out to verify the system performance.

#### 2) Usage of Eyeblink to control status (Go/Stop) of the wheelchair

We have designed a system in which the subject blinks thrice in a 4s window to start (go) the wheelchair, and the same mechanism is used to stop the wheelchair from moving. The subjects were asked to blink consecutively thrice in a period of 4s during which the data was recorded from Fp1, Fp2 electrodes, and commands were generated. 12 trials of blinking were carried out during the online session for each subject. The eye blink detection ran in parallel with the SSVEP classification. Fig. 1 shows the schematic diagram of the wheelchair prototype design based on a combination of SSVEP and eyeblinks.

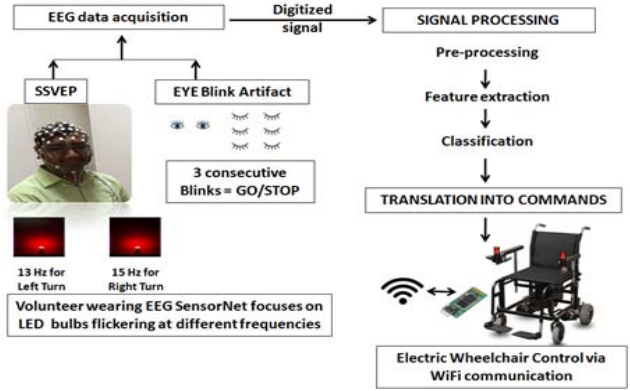

Figure 1. Schematic diagram of the proposed SSVEP and eye blink based electric wheelchair control technology

C. Algorithms

*1) Eyeblink analysis*

The data was bandpass filtered (0.1-70Hz), segmented and then converted into MATLAB file format for further processing. A MATLAB-based script was designed to detect voluntary blinks. It underwent pattern matching with an ideal eye blink data, which was present inherently in the program. The program did pattern matching between the recorded EEG data and the ideal eye blink using the principle of cross-correlation. The cross-correlation values (CCV) were calculated simultaneously. If the ideal eye blink encountered an eye blink artifact in the fed EEG data during the pattern matching process, it recorded a high CCV. Then, a threshold value was assigned to the CCV to identify the eye blink artifact in the EEG data as an actual eye blink. The source code for this software program was inspired by a set of eye blink applications on BCI called icablinkmetrices3.1 [24]. The eye blink data analysis resulted in perfect 3 peak detection as per the designed protocol. These markers were sent to the RaspberryPi to generate commands to start (go) and stop the wheelchair.

*2) SSVEP analysis*

The experiment was conducted by flickering two LEDs at different frequencies. From the real-world application point of view, the user needs to control the wheelchair in 2 primary directions: left and right. To achieve that, we have placed 2 flickering LEDs, 13Hz (for left) & 15 Hz (for right), on the armrest of the wheelchair. The subject focused on a particular LED for 4s to choose the desired direction. As the two LEDs flickered simultaneously, the SSVEP signals were found to be interfering with each other. Therefore, we used machine learning (ML) to predict the user's intention accurately.

*3) Data preprocessing: Wavelet packet denoising*

The raw SSVEP data was segmented using Netstation Tools. It was referenced and detrended to give a zero mean. Then 15s segments of each trial were obtained. Moving windows of 4s each were taken for analysis, with the sampling rate being 1000ms. Mostly it is seen that SSVEP data contains a linear combination of signals from various sources, which also includes unanticipated muscle and eye movements [25]. The filtering of such mixtures through regular filters is ineffective. However, the wavelet-based denoising performs a correlation analysis. Therefore, the output is expected to be maximal when the input signal resembles maximum to the mother wavelet. In our study, a wavelet packet tree was generated by applying the wavelet packet transform to the noisy signal. The mother wavelet used for decomposition was sym9 [26]. Then threshold selection by Stein's Unbiased Estimate of Risk [27-29] was performed. The risk was estimated and further minimized to obtain the optimized threshold value. The threshold was applied to the coefficients signals using the thresholding filter selected, and modified coefficients were obtained [30,31]. Finally, inverse wavelet packet transform was used to get the threshold coefficients, and the denoised signal was obtained.

*4) Feature extraction using WPD and CCA*

Wavelet packet decomposition (WPD) is an extension of the wavelet transform and used in many research work [32, 33]. Methodically, the wavelet decomposition splits the original signal into two subspaces, V and W. V being the space that includes the low-frequency information about the original signal, and W includes the high-frequency information. The V and W are again split in the same way, and the process is repeated as per the required frequency band. As db7 exhibited the highest similarities and compatibility for EEG signals in the occipital regions [34], the signal from each electrode (O1, O2, Oz) was thereby decomposed into 8 levels using the db7 wavelet. Subsequently, the node corresponding to the relevant frequency band of interest was reconstructed. The 3 nodes, each corresponding to frequency band containing the stimulation frequency and its 2 harmonics was reconstructed in the following way: (64) for F1Hz & F2Hz, (66) for 2F1Hz & 2F2Hz and (68) for 3F1Hz & 3F2Hz, respectively. Henceforth, 9 signals were extracted for each segment corresponding to 3 nodes for 3 electrodes each.

Canonical correlational analysis (CCA) is the state-of-the-art method used to translate the users' EEG signals into the corresponding frequency of the flickering stimuli. In our work, CCA coefficients for a 4s window were calculated for the signal for sine and cosine waves of F1, 2×F1, 3×F1, F2, 2×F2, 3×F2. The CCA coefficient was calculated between the reconstructed signal from WPD and the appropriate harmonic of the sine-cosine wave of simulating frequency, i.e., for each reconstructed node. The coefficient was obtained for nF1 and nF2 (n:1,2,3). Therefore, for each of the 9 signals obtained from the wavelet reconstruction, 2 CCA coefficients were calculated, and hence 18 features were extracted for each segment of EEG time series data.

*5) Classification*

In order to deal with multi-class classification, we adopted "one against one" SVM, which constructs k (k-1)/2 (where k is the number of classes and k is 3 in this study) classifiers where each one is trained on data from two classes [35]. In our work, the training process was conducted with 5 folds cross-validation to prevent over-fitting. For decision making, segments of different sizes (2s,3s,4s,5s,6s) of EEG time series signal were considered, and a 4s window was chosen, which was found to be optimum for our experiment. The selection of this window length was based on optimal accuracy and speed [36].

Cross-validation during training improves the accuracy of the model by avoiding overfitting [37]. In K-fold cross-validation, the data is first partitioned into K equally (or nearly equally) sized segments or folds. For each of K folds, we use K−1 folds for training and the remaining one for testing the hyperparameter (Gamma and C). The gamma parameters can be seen as the inverse of the radius of influence of samples selected by the model as support vectors. C behaves as a regularization parameter in the SVM [38]. In this work, 5-fold cross-validation was used to optimize the hyperparameters of the SVM and estimate model performance. Herein the selection of optimal hyperparameters was made using Bayesian optimization [39, 40], and after hyperparameter tuning, the resulting final model was used for testing.

### III. RESULTS AND DISCUSSION

The training and testing accuracy of all the subjects is presented in Table 2. It was observed that Subject-1 had the highest accuracy both in terms of cross-validation and test, while Subject-11 showed the minimal value in the same. The average cross-validation accuracy was found to be 89.65%, and the test accuracy was averaged out to be 83.53%. None of the participants complained about any visual strain and fatigue during the task. In our work, we extracted the CCA coefficients for narrowband signals reconstructed from the wavelet packet decomposition, which is not reported elsewhere in the authors' knowledge.

TABLE II
ACCURACY (SUBJECT WISE TRAINING AND TESTING) WITH THE SVM METHOD.

| Subject No. | Accuracy (Cross Validation) (%) | Accuracy (Test) (%) |
|---|---|---|
| 1 | 99.33 | 96.98 |
| 2 | 89.05 | 91.24 |
| 3 | 93.97 | 88.76 |
| 4 | 93.87 | 93.46 |
| 5 | 91.66 | 86.58 |
| 6 | 94.05 | 84.90 |
| 7 | 86.86 | 82.86 |
| 8 | 80.81 | 79.53 |
| 9 | 96.80 | 74.33 |
| 10 | 89.36 | 72.15 |
| 11 | 78.76 | 69.30 |
| 12 | 81.26 | 82.55 |
| Average | 89.65 | 83.53 |
| Standard Deviation | 6.60 | 8.59 |
| Final | 89.65 ± 6.6 | 83.53 ± 8.59 |

#### A. Wheelchair interfacing with EEG signals

The electric wheelchair's (Ostrich Mobility Instruments Private LTD., India) joystick controls were incorporated in the BCI system using a mechanical contraption designed in-house. The wheelchair was connected through a WIFI network using HTTP requests through specific programs provided by EGI (AmpServer SDK) that collects EEG data from the subject. The block diagram shown in Fig.2 includes all the elements required to control the wheelchair starting from the computer interface to the RaspberryPi microcontroller interface to the motors responsible for moving the wheelchair.

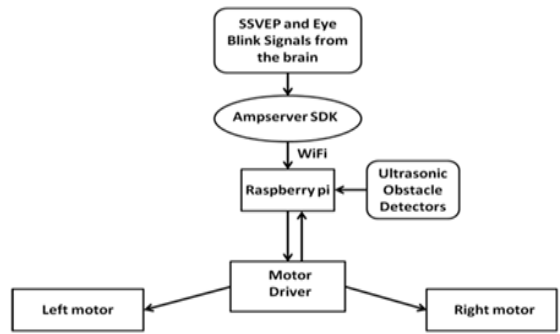

Figure 2: The core unit of the hardware system is the RaspberryPi microcontroller

Fig. 2 depicts the flow logic of the driving mechanism of it. The commands generated as Left, Right, Go and Stop were transferred to the joystick hub via WIFI using an HTTP Post request parked by the RaspberryPi Python script to drive the stepper motors on the wheelchair joystick contraption. The actual testing of the wheelchair automation was carried out with 12 healthy subjects for intended vs. actual L (left), R (right) turns, and GO/STOP commands (Table 3).

TABLE III
FINAL TESTING OF WHEELCHAIR AUTOMATION BY SSVEP AND EYE BLINK COMMANDS.

| Subject No. | Intended L turn | Actual L turn | Intended R turn | Actual R turn | Intended GO/STOP | Actual GO/STOP | Success Rate (%) | ITR (bits/min) |
|---|---|---|---|---|---|---|---|---|
| 1 | 10 | 10 | 10 | 9 | 12 | 12 | 96.87 | 69.40 |
| 2 | 10 | 8 | 10 | 9 | 12 | 11 | 87.5 | 57.34 |
| 3 | 10 | 9 | 10 | 10 | 12 | 12 | 96.87 | 69.4 |
| 4 | 10 | 10 | 10 | 8 | 12 | 10 | 87.5 | 57.34 |
| 5 | 10 | 8 | 10 | 8 | 12 | 10 | 81.25 | 50.43 |
| 6 | 10 | 9 | 10 | 9 | 12 | 10 | 90.62 | 61.06 |
| 7 | 10 | 8 | 10 | 8 | 12 | 10 | 81.25 | 50.43 |
| 8 | 10 | 8 | 10 | 8 | 12 | 11 | 84.37 | 53.80 |
| 9 | 10 | 8 | 10 | 8 | 12 | 10 | 81.25 | 50.43 |
| 10 | 10 | 9 | 10 | 7 | 12 | 10 | 81.25 | 50.43 |
| 11 | 10 | 8 | 10 | 9 | 12 | 10 | 84.37 | 53.80 |
| 12 | 10 | 9 | 10 | 9 | 12 | 11 | 90.62 | 61.06 |
| AVG | | | | | | | 86.97 | 56.71 |

The ITR was calculated using the formula stated by Wolpaw et al. [41]. For $n$ trials, identification accuracy $p$, and time for a command $t$,

$$ITR \left(\frac{bits}{min}\right) = \frac{\left(\log n + p \log p + (1-p) \log \frac{1-p}{n-1}\right) 60}{t}$$

The average ITR was found out to be 56.71 bits/min for 12 subjects (Table 3). The time taken (t) for each command execution was 4.015s for the left or rightward turn at about 0-60° rotation. Each action automatically stopped until it received the next command.

### B. Safety feature and emergency force stop

In order to provide obstacle detection at about 0.5m in all three directions, the prototype was additionally assembled with three ultrasonic sensors (HC-SR04; range: 2-400cm, 30°), one at the front and two above the front wheels. These ultrasonic sensors were programmed by RaspberryPi (Fig.2). While moving when the wheelchair was about to strike any obstacle at about 0.5m in any direction, the sensors detected it, and a beep or alarm sound was generated. Instantly the wheelchair was force stopped until the following start command.

### C. Novelty and key contributions

In our prototype design, using the tiny LED bulbs (3mm) mounted on the armrest of the wheelchair, there was no front vision block for the subjects. These LEDs were more convenient as they accounted for only left and right direction control and needed to be stared at for a maximum of 4s. Likewise, the eye blinks were relatively easy to perform and were effectively used for STOP/GO toggle for the wheelchair. The novelty of this prototype is a simple combination of SSVEP and eye blink commands to control the wheelchair with a reasonable level of accuracy and without causing any discomfort to the user. This also demands significantly less training time for the subjects.

Additionally, it is equipped with the simplest form of obstacle detecting sensors and an emergency force stop system with its advantage. The SSVEP signal detection pipeline has new and improved features such as the combination of Wavelet packet denoising, CCA over narrowband wavelet packets, and SVM with Bayesian optimization that offers high accuracy over 12 subjects. The same has the potential to be effectively used for other BCI applications.

## IV. CONCLUSION

This wheelchair prototype presents a convenient and efficient user interface combining SSVEP and eye blinks with an average of 83.53% accuracy of the classifying system. In all the trials, the subjects were able to control the wheelchair quite comfortably and as intended by simple gazing and blinking with an average success rate of 86.97%. For each command execution, the time taken was 4.015s which is considered nominal and apt for the intended task. The results showed the performance of the hybrid BCI with comparable accuracy and ITR. In addition, obstacle detection using ultrasonic sensors and a force stop system offers basic safety to the user while driving. In the future task, it is planned to test this wheelchair prototype by involving some actual patients. The safety features would also be improvised in the next plan of action, for e.g., to pass through a doorway; evade obstacles effortlessly, among other tasks. In the future, the system could also be integrated with various techniques by modularization. The proposed wheelchair prototype is expected to provide a pathway for re-establishing convenient, safe, and comfortable communication for people with severe disabilities without being dependent on others. Promisingly, this setup can be implemented as a proof of concept for other home automation-based experiments.

REFERENCES

[1] K. R. Müller, B. Blankertz, "Toward Non-invasive Braincomputer Interfaces," IEEE Signal Processing Magazine, vol. 128, no. 1, pp.125-128, 2006.

[2] T. W. Berger, J. K. Chapin, G. A. Gerhardt, D. J. McFarland, J. C. Principe, W. V. Soussou, D. M. Taylor, P. A. Tresco, "International assessment of research and development in brain-computer interfaces: report of the world technology evaluation center," Berlin: Springer, 2007.

[3] M. Cheng, X. R. Gao, S. K. Gao, and D. Xu, "Design and implementation of a brain computer interface with high transfer rates," IEEE Transactions on Biomedical Engineering, vol. 49. No. 10, pp. 1181-1186, 2002.

[4] G. R. Muller-Putz, R. Scherer, C. Brauneis, and G. Pfurtscheller, "Steady-State Visual Evoked Potential (SSVEP)-based Communication: impact of harmonic frequency components," Journal of Neural Engineering, vol. 2, no. 4, pp. 123-130, 2005.

[5] G. R. Muller-Putz and G. Pfurtscheller, "Control of an electrical prosthesis with an SSVEP-based BCI," IEEE Transactions on Biomedical Engineering, vol. 55, no. 1, pp. 361–364, 2008.

[6] Marcin Jukiewicz, Anna Cysewska-Sobusiak, "Stimuli design for SSVEP-based brain computer-interface", International Journal of Electronics and Telecommunication, vol. 62, no. 2, pp. 109–113, 2016.

[7] B. Rebsamen, E. Burdet, C. Guan, C. L. Teol, Q. Zeng, M. Ang, C. Laugier, "Controlling a wheelchair using a BCI with low information transfer rate," 2007 IEEE 10th International Conference on Rehabilitation Robotics, Noordwijk, 2007, pp. 1003-1008. doi: 10.1109/ICORR.2007.4428546.

[8] Y. Li, J. Pan, F. Wang and Z. Yu, "A Hybrid BCI System Combining P300 and SSVEP and Its Application to Wheelchair Control," in IEEE Transactions on Biomedical Engineering, vol. 60, no. 11, pp. 3156-3166, Nov. 2013. doi: 10.1109/TBME.2013.2270283.

[9] N. Mora, I. De Munari, P. Ciampolini (2015) Improving BCI Usability as HCI in Ambient Assisted Living System Control. In: Schmorrow D., Fidopiastis C. (eds) Foundations of Augmented Cognition. AC 2015. Lecture Notes in Computer Science, vol 9183. Springer, Cham.

[10] A. A. Ghodake and S. D. Shelke, "Brain controlled home automation system," 10th International Conference on Intelligent Systems and Control (ISCO), Coimbatore, 2016, pp.1-4. doi: 10.1109/ISCO.2016.7727050.

[11] P. F. Diez, S. M. T. Müller, V. A. Mut, E. Laciar, E. Avila, T. F. Bastos-Filho, M. S. Filho, "Commanding a robotic wheelchair with a high-frequency steady-state visual evoked potential based brain-computer interface". Medical Engineering & Physics, vol. 35(8), pp. 1155–1164, 2013.

[12] J. Duan, Z. Li, C. Yang, P. Xu, "Shared control of a brain- actuated intelligent wheelchair", 11th World Congress on Intelligent Control and Automation (WCICA), 2014, pp 341–6.

[13] D. W-K. Ng, Y-W. Soh, S-Y Goh, "Development of an autonomous BCI wheelchair". IEEE Symposium on Computational Intelligence in Brain ComputerInterfaces (CIBCI) 2014 Dec., pp. 1-4.

[14] S. M. T. Müller, T. F. Bastos, M. S. Filho, "Proposal of a SSVEP-BCI to command a robotic wheelchair". Journal of Control, Automation and Electrical Systems, vol. 24, pp. 97–105, 2013.

[15] Z. Xu, J. Li, R. Gu, B. Xia, "Steady-state visually evoked potential (SSVEP)-based brain-computer interface (BCI): a low-delayed asynchronous wheelchair control system". Neural Information Processing. Iconip 2012, Part I, vol. 7663, pp. 305–14. Available from: <Go to ISI>:// WOS:000345086000037.

[16] C. Mandel, T. Lüth, T. Laue, T. Röfer, A. Gräser, B. K. Brückner, "Navigating a smart wheelchair with a brain-computer interface interpreting steady-state visual evoked potentials", IEEE/RSJ International Conference on Intelligent Robotics Systems. St. Louis, MO, 2009, pp. 1118-1125.doi: 10.1109/IROS.2009.5354534.

[17] F-B. Vialatte, M. Maurice, J. Dauwels, A.Cichocki, "Steady-state visually evoked potentials: focus on essential paradigms and future perspectives". Progress in Neurobiology, vol. 90, pp. 418-438, 2015.

[18] Y. Wang, R. Wang, X. Gao, B. Hong, S. Gao, "A practical VEP-based brain-computer interface". IEEE Transactions on Neural System and Rehabilitation Engineering, vol.14, pp. 234-239, 2006.

[19] G. Bin, X. Gao, Z. Yan, B. Hong, S. Gao, "An online multi-channel SSVEP-based brain–computer interface using a canonical correlation analysis method". Journal of Neural Engineering, vol. 6, pp. 46002, 2009.

[20] J. Parra, F. H. L. da Silva, H. Stroink, S. Kalitzin, "Is colour modulation an independent factor in human visual photosensitivity?", Brain, vol. 130(6), pp. 1679–1689,June 2007 https://doi.org/10.1093/brain/awm103.

[21] D. Zhu, J. Bieger, G. G. Molina, and R. M. Aarts, "A survey of simulation methods used in SSVEP-based BCIs". Intell. Neuroscience 2010, Article 1 (January 2010), 12 pages. DOI:http://dx.doi.org/10.1155/2010/702357.

[22] T. Cao, F. Wan, P. U. Mak, P. Mak, M. I. Vai and Y. Hu, "Flashing color on the performance of SSVEP-based brain-computer interfaces," 2012 Annual International Conference of the IEEE Engineering in Medicine and Biology Society, San Diego, CA, 2012, pp. 1819-1822. doi: 10.1109/EMBC.2012.6346304.

[23] R.M.G. Tello, S.M.T. Müller, T. Bastos-Filho, A. Ferreira, "A comparison of techniques and technologies for SSVEP classification", Biosignals and Biorobotics Conference, Biosignals and Robotics for Better and Safer Living (BRC), 5th ISSNIP-IEEE (2014), pp. 1-6.

[24] M. B. Pontifex, V. Miskovic, S. Laszlo, "Evaluating the efficacy of fully automated approaches for the selection of eyeblink ICA components", Psychophysiology, vol.54(5),pp.780-791, 2017. doi: 10.1111/psyp.12827.

[25] J. Wu, J. Zhang, L. Yao, "An automated detection and correction method of EOG artifacts in EEG-based BCI", ICME International Conference on Complex Medical Engineering, Tempe, AZ, pp. 1-5, 2009doi: 10.1109/ICCME.2009.4906624.

[26] N. K. Al-Qazzaz, B. M. A. S. Hamid, S. A. Ahmad, M. S. Islam, J. Escudero, "Selection of mother wavelet functions for multi-channel eeg signal analysis during a working memory task". Sensors, vol.15, pp. 29015–29035, 2015.

[27] C. M. Stein, "Estimation of the mean of a multivariate normal distribution". Annals of Statistics, vol. 9, pp. 1135–1151, 1981.

[28] R. Romo-Vazquez, R. Ranta, V. Louis-Dorr, D. Maquin, "EEG ocular artefacts and noise removal". In Proceedings of the 29th Annual International Conference of the IEEE Engineering in Medicine and Biology Society, Lyon, France, 22–26 August 2007; pp. 5445–5448.

[29] M. Sheoran, S. Kumar, and A. Kumar, "Wavelet-ICA based de-noising of electroencephalogram signal," International Journal of Information and Computation Technology, vol. 4, pp. 1205-1210, 2014.

[30] M. Patil, N. Garg, L. Kanungo and V. Baths, "Study of motor imagery for multiclass brain system interface with a special focus in the same limb movement," 2019 IEEE 18th International Conference on Cognitive Informatics & Cognitive Computing (ICCI*CC), 2019, pp. 90-96, doi: 10.1109/ICCICC46617.2019.9146105.

[31] A. Valsaraj, I. Madala, N. Garg, M. Patil and V. Baths, "Motor Imagery Based Multimodal Biometric User Authentication System Using EEG," 2020 International Conference on Cyberworlds (CW), 2020, pp. 272-279, doi: 10.1109/CW49994.2020.00050.

[32] L. Zhao, H. Shen, Y. Bian, L. Xiao, D. Hu, P. Yuan, "Application of wavelet packet technique in BCI," 2009 IEEE International Conference on Intelligent Computing and Intelligent Systems, Shanghai, 2009, pp. 43-46.doi: 10.1109/ICICISYS.2009.5358230.

[33] Z. Lin, C. Zhang, W. Wu, X. Gao. "Frequency recognition based on canonical correlation analysis for SSVEP-based BCIs". IEEE Transactions in Biomedical Engineering, vol.54, pp. 1172–1176, 2007.

[34] J. Pan, X. Gao, F. Duan, Z. Yan, S. Gao, "Enhancing the classification accuracy of steady-state visual evoked potential-based brain–computer interfaces using phase constrained canonical correlation analysis", Journal of Neural Engineering, vol. 8(3), pp. 036027, 2011.

[35] Y. Zhang, G. Zhou, J. Jin, X. Wang, A. Cichocki, "Frequency recognition in SSVEP-based BCI using multiset canonical correlation analysis", International Journal of Neural Systems, vol. 24(4), pp. 1-7, 2013.

[36] JH. Jian and K. Tang, "Improving classification accuracy of SSVEP based BCI using RBF SVM with signal quality evaluation," 2014 International Symposium on Intelligent Signal Processing and Communication Systems (ISPACS), Kuching, 2014, pp. 302-306. doi: 10.1109/ISPACS.2014.7024473.

[37] R. O. Duda, P. E. Hart, D. G. Stork, "Pattern classification", 2nd ed. New York: John Wiley & Sons. 2000.

[38] C. Cortes, V. Vapnik, "Support-vector network". Machine Learning, vol. 20, pp. 273–297, 1995.

[39] J. Mockus, V. Tiesis, A. Zilinskas, "The application of Bayesian methods for seeking the extremum" in Toward Global Optimization, Elsevier: Amsterdam, The Netherlands, vol. 2, pp. 117–129,1978.

[40] C. E. Rasmussen, C. Williams, "Gaussian Processes for Machine Learning", MIT Press, 2006.

[41] R. Wolpaw, D. J. McFarland, T. M. Vaughan, "Brain-computer interface research at the Wadsworth Center", IEEE Transactions on Neural Systems and Rehabilitation Engineering. Vol.8(2), pp. 222–226, 2000.